# The meaning behind observed $p_T$ regions at the LHC energies

Mais Suleymanov

Department of Physics, COMSATS Institute of Information Technology, Islamabad Pakistan

We argue that $p_T$ distribution data from the LHC on the invariant differential yield of the charged primary particles in $pp$ collisions at $\sqrt{s_{NN}} = 0.9\,\text{TeV}, 2.76\,\text{TeV}, 7\,\text{TeV}$ and in Pb-Pb collisions at $\sqrt{s_{NN}} = 2.76\,\text{TeV}$ with 6 centrality bins contains several $p_T$ regions with special properties. These distributions were analysed by fitting the data with exponential functions. We conclude that the regions reflect features of fragmentation and hadronization of partons through the string dynamics. The nuclear transparency results in negligible influence of the medium in the *III* region ($p_T > 17 - 20\,\text{GeV}/\text{c}$), which has highest $p_T$ values. The effects and changes by the medium start to appear weakly in the *II* region ($4 - 6\,\text{GeV}/\text{c} < p_T < 17 - 20\,\text{GeV}/\text{c}$) and become stronger in the *I* region ($p_T < 4 - 6\,\text{GeV}/\text{c}$). It seems that the *II* region has highest number of strings. The increase in string density in this region could lead to fusion of strings, appearance of a new string and collective behaviour of the partons in the most central collisions. These phenomena can explain anomalous behaviour of the Nuclear Modification Factor in the *II* region. We propose the *II* region as a possible area of Quark Gluon Plasma formation through string fusion. The first $p_T$ regions are the ones with the maximum number of hadrons and minimum number of strings due to direct hadronization of the low energy strings into two quark systems - mesons.

## I. INTRODUCTION

The work in the article [1] presents the evolution picture of the Nuclear Modification Factor with center-of-mass energy ($\sqrt{s_{NN}}$) from the SPS [2] to the RHIC [3] and then to the LHC [4]. The creation of high-$p_T$ particles in central Pb-Pb collisions is significantly suppressed in comparison to peripheral Pb-Pb ($R_{CP}$) and $pp$ collisions ($R_{AA}$). In the range of $p_T = 5 - 10\,\text{GeV}/\text{c}$, the suppression is stronger than before observed at the RHIC [3]. Beyond $p_T = 10\,\text{GeV}/\text{c}$ up to 20 GeV/c both $R_{CP}$ and $R_{AA}$ show a rising trend as it was shown by the data from the ALICE experiment [4]. The CMS measurement [1] with improved statistical precision clearly shows that this rise continues at higher $p_T$, approaching a suppression factor $R_{AA} \approx 0.5 - 0.6$ in the range of 40–100 GeV/c. The behaviour of $R_{CP}$ and $R_{AA}$ as a function of $p_T$ are very complex and today we are far from understanding they exactly. However, the overall $p_T$ dependence of the suppression can be described by a number of phenomenological predictions, and the detailed evolution of the $R_{AA}$ increasing trend from 6 to 100 GeV/c depends on the input parameters of the model (see in the paper [1]).

This paper presents the results of analysis of invariant differential yield of the primary charged particle as a function of $p_T$ in *pp* collisions at $\sqrt{s} = 0.9\,\text{TeV}$ [5], 2.76 TeV [1] and 7 TeV [5] and in Pb-Pb collisions at $\sqrt{s_{NN}} = 2.76\,\text{TeV}$ with 6 centrality bins [1] which can help to understand the behaviour of $R_{CP}$ and $R_{AA}$ as a function of $p_T$. We derived the results from fitting the $p_T$ distributions with the exponential function. In section II we show that the $p_T$ distributions contain some regions for *pp* and Pb-Pb collisions with different centrality using our fitting results and results presented by other groups. Section III describes a more detailed study of the characteristics of the $p_T$ regions using the fitted data, including the lengths of the regions and behaviors of the fitting parameters. The main results are presented in section IV, and their discussions and conclusions are given in sections V and VI, respectively.

## II. OBSERVATION OF THE $p_T$ REGIONS

Figure 1 shows $p_T$ distributions of the invariant charged particles differential yield averaged over the pseudorapidity $|\eta|<1$ for charged primary particles produced in *pp* collisions (index *c=1* specifies *pp* collisions) and Pb-Pb interactions with different centralities (c=21,22,23,24,25,26 are used to designate Pb-Pb collisions with centrality bins: 0÷5%, 5÷10%, 10÷30%, 30÷50%, 50÷70%, 70÷90% respectively) at 2.76 A TeV [1][1]. We observed at least three $p_T$ regions with different behaviours. Figure 2 shows $p_T$ distributions for the invariant charged particle differential yield (averaged over the pseudorapidity $|\eta|<2.4$) in *pp* collisions at $\sqrt{s} = 0.9\,\text{TeV}$ (*c=3*) and $\sqrt{s} = 7\,\text{TeV}$ (c=4) [5][2]. Here, we again saw several qualitatively different $p_T$ regions[3].

First, we fit[4] the $p_T$ distributions in the whole experimentally measured ranges with a simple function $y = ae^{-bx}$, where *a* and *b* are free parametes[5], ignoring the observed deviations in behaviour for different regions. This attempt did not produce good fitting results. There are two different ways to improve fitting. The first approach is to apply a more complex fitting function, e.g. Tsallis distribution [8], and increase the number of free parameters. The second one is to fit the original simple function in the limited regions of $p_T$. We have taken the second approach motivated by the idea of the existence of different regions in the $p_T$ distributions. To get the best fitting results we varied the values of the $p_T$ from $p_T^{min}$ to $p_T^{max}$ and fitted separate regions with the original distribution. The best results of the fitting are shown in Table I. From the data presented in the table, one can see that there are three distinct regions of $p_T$ distributions of the charged particles produced in $c=1,3, 21-26$ collisions and four regions in the case of $c=4$. So one can conclude that the fitting results confirm the existence of several $p_T$ regions for the behaviour of the invariant charged particle differential yield as a function of $p_T$ in $pp$ collisions and Pb-Pb collisions with different centrality bins.

---

[1] The experimental data were taken from the HEP Data: https://hepdata.net/record/ins1088823

[2] The experimental data were taken from the HEP Data: https://hepdata.net/record/ins896764

[3] In the paper [6] the authors noted that in the measured charged-particle $p_T$ spectrum, which covers the range $p_T < 20\,\text{TeV}$ (first ALICE physics results from the $\sqrt{s_{NN}} = 2.76\,\text{TeV}$ Pb-Pb collisions [7]) and extends over three orders of magnitude, two quite distinct regions could be seen for the central most collisions: at $p_T \leq 4-5\,\text{GeV}$ the spectrum is exponential, while at $p_T \geq 5\,\text{GeV/c}$ it shows a power-law like behaviour.

[4] To fit the distributions we used ROOT soft (version 5.34/02, 21$^{st}$ September 2012), taking into account the statistical and systematic uncertainties of the measurement.

[5] In the paper [8] the $p_T$ distribution of the particles produced in the *pp* collisions (data from the STAR, PHENIX, ALICE and CMS) were fitted using the Tsallis distribution. They concluded that there exists a universal distribution for fitting. The distribution has 3 free parameters. However the Tsallis distribution can only fit part of the particle spectra in Pb-Pb collisions at $\sqrt{s_{NN}} = 2.76\,\text{TeV}$ [8], either in the low or high p$_T$ region. The authors [8] have proposed a new formula in order to fit all the particle spectra in Pb-Pb by increasing one fitting degree of freedom from Tsallis distribution. They noted that this follows the same idea of the transition from the exponential distribution to Tsallis distribution when intermediate $p_T$ data are available in experiments.

Before moving on to study the characteristics of the extracted regions let us consider some other indications that the existence of qualitatively different regions is a fact and not an artefact of the fitting method.

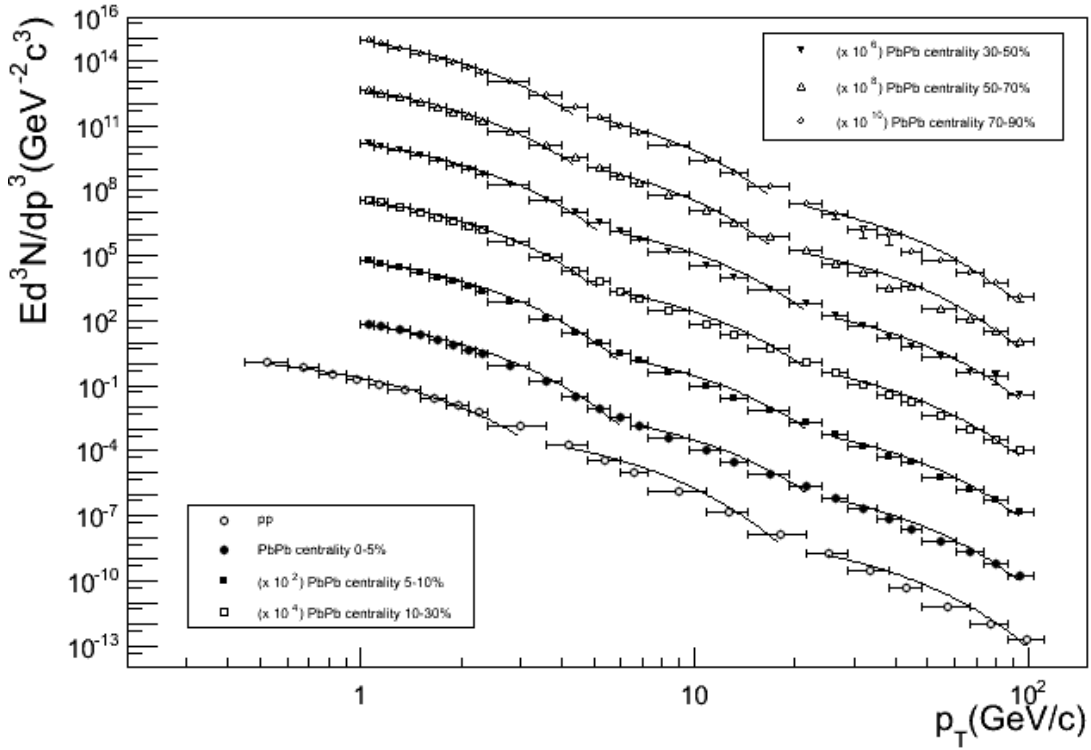

FIG. 1. Invariant charged particle differential yield for $|\eta|<1$ in Pb-Pb collisions and in $pp$ collisions at $\sqrt{s_{NN}} = 2.76\,\text{TeV}$ [1]. The lines in the figure show the fitting results.

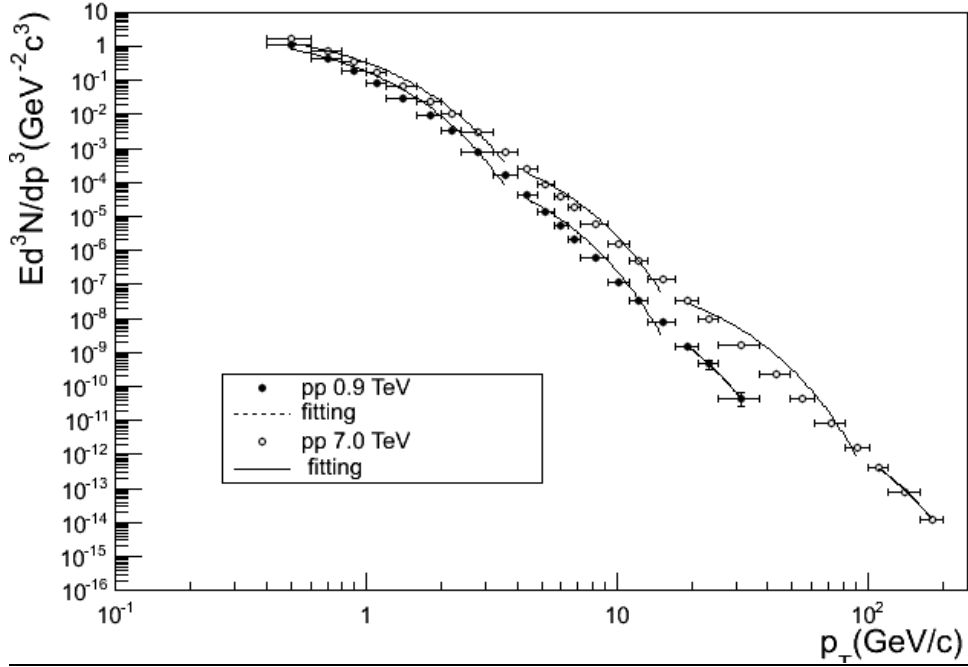

FIG. 2. The $p_T$ distributions for the invariant charged particle differential yield (averaged over the pseudorapidity $|\eta| < 2.4$) in $pp$ collisions at $\sqrt{s} = 0.9\,\text{TeV}$ and $\sqrt{s} = 7\,\text{TeV}$ [5].

TABLE I.

| c | K → | I | II | III |
|---|---|---|---|---|
| 3 | $p_T^{min} \div p_T^{max}$ (GeV/c) | (0.5±0.1) ÷ (3.6±0.4) | (4.4±0.4)÷(15.2±2.0) | (19.2±2.0) ÷ (31.2±6.0) |
|   | $J^c$ | $J^c_{I-II}$ : 4.0±0.3 | $J^c_{II-III}$ :17.2±1.4 | $J^c_{III-IV}$ : 31.2±6.0 |
|   | $L^c_K$ | 4.0±0.3 | 13.2±1.4 | 14.0±6.2 |
|   | $\chi^2/ndf$; Prob. | 5.972/7;0.5431 | 5.805/6;0.4454 | 0.0005462/1; 0.9814 |
|   | $a$(GeV$^{-2}$c$^3$) | 3.8±0.8 | 0.0014±0.0005 | (4.2±12.7) 10$^{-7}$ |
|   | $b$ (GeV/c)$^{-1}$ | 3.0±0.1 | 0.87 ± 0.05 | 0.29±0.14 |
| 4 | $p_T^{min} \div p_T^{max}$ (GeV/c) | (0.5±0.1)÷(3.6±0.4) | (4.4±0.4)-(15.2±2.0) | (19.2±2.0)÷(91.2±10.0) |
|   | $J^c$ (GeV/c) | $J^c_{I-II}$ : (4.0±0.3) | $J^c_{II-III}$ : (17.2±1.4) | $J^c_{III-IV}$ : (101.1±7.1) |
|   | $L^c_K$ (GeV/c) | (4.0±0.3) | (13.2±1.4) | (83.9±7.2) |
|   | $\chi^2/ndf$; Prob. | 6.55/7;0.4771 | 5.632/6;0.4656 | 4.238/5;0.5157 |
|   | $a$(GeV$^{-2}$c$^3$) | 4.8±0.9 | 0.005±0.002 | (4.5±1.5) 10$^{-7}$ |
|   | $b$ (GeV/c)$^{-1}$ | 2.7±0.1 | 0.76±0.04 | 0.146±0.009 |

| | | | | IV Region |
|---|---|---|---|---|
| | $p_T^{\min} \div p_T^{max}$ (GeV/c) | | | (111±10)÷(181±20) |
| | $J^c$ (GeV/c) | | | $J_{IV-V}^c$ : (181±20) |
| | $L_J^c$ (GeV/c) | | | (79.9±21.2) |
| | $\chi^2/ndf$; Prob. | | | 0.007805/1;0.9296 |
| | $a$(GeV$^{-2}$c$^3$) | | | (0.95±1.9) 10$^{-10}$ |
| | $b$ (GeV/c)$^{-1}$ | | | 0.049±0.015 |
| 1 | $p_T^{\min} \div p_T^{max}$ (GeV/c) | (0.53±0.08) ÷(3.0±0.6) | (4.2±0.6)÷(18.0±3.6) | (25.2±3.6)÷(99.3±12.9) |
| | $J^c$ (GeV/c) | $J_{I-II}^c$ : (3.6±0.4) | $J_{II-III}^c$ : (21.6±2.5) | $J_{III-IV}^c$ : (100.0±12.9) |
| | $L_J^c$ (GeV/c) | (3.6±0.4) | (18.0±2.5) | (78.4±13.3) |
| | $\chi^2/ndf$; Prob. | 4.132/8;0.845 | 3.458/4; 0.4842 | 2.47/4;0.65 |
| | $a$(GeV$^{-2}$c$^3$) | 5.16±0.96 | 0.0030±0.0014 | (4.0±2.3) 10$^{-5}$ |
| | $b$ (GeV/c)$^{-1}$ | 3.1±0.2 | 0.74±0.06 | 0.13±0.01 |
| 21 | $p_T^{\min} \div p_T^{max}$ (GeV/c) | (1.05±0.05)÷(6.0±0.4) | (6.8±0.4)÷(21.6±2.4) | (26.4±2.4) ÷ (95.0±8.6) |
| | $J^c$ (GeV/c) | $J_{I-II}^c$ : (6.4±0.3) | $J_{II-III}^c$ : (24±1.7) | $J_{III-IV}^c$ : (100.0±8.6) |
| | $L_J^c$ (GeV/c) | (6.4±0.3) | (17.6±1.7) | (76.0±8.8) |
| | $\chi^2/ndf$; Prob. | 11.24/11;0.4232 | 2.678/4; 0.613 | 4.447/6;0.6164 |
| | $a$(GeV$^{-2}$c$^3$) | 631±71 | 0.037±0.014 | (1.6±0.6) 10$^{-5}$ |
| | $b$ (GeV/c)$^{-1}$ | 2.18±0.06 | 0.48±0.04 | 0.128±0.007 |
| 22 x10$^2$ | $p_T^{\min} \div p_T^{max}$ (GeV/c) | (1.05±0.05) ÷ (6.0±0.4) | (6.8±0.4)÷(21.6±2.4) | (26.4±2.4) ÷(95.0±8.6) |
| | $J^c$ (GeV/c) | $J_{I-II}^c$ : (6.4±0.3) | $J_{II-III}^c$ : (24±1.7) | $J_{III-IV}^c$ : (100.0±8.6) |
| | $L_K^c$ (GeV/c) | (6.4±0.3) | (17.6±1.7) | (76.0±8.8) |
| | $\chi^2/ndf$; Prob. | 12.14/11;0.3535 | 2.697/4;0.6098 | 3.331/6; 0.7663 |
| | $a$(GeV$^{-2}$c$^3$) | (5.0±0.6)10$^5$ | 36±14 | 0.011±0.004 |
| | $b$ (GeV/c)$^{-1}$ | 2.15±0.06 | 0.48±0.04 | 0.125±0.008 |
| 23 x10$^4$ | $p_T^{\min} \div p_T^{max}$ (GeV/c) | (1.05±0.05) ÷(5.2±0.4) | (6. ±0.4)÷(21.6±2.4) | (26.4±2.4) ÷(95.0±8.6) |
| | $J^c$ (GeV/c) | $J_{I-II}^c$ : (5.6±0.3) | $J_{II-III}^c$ : (24±1.7) | $J_{III-IV}^c$ : (100.0±8.6) |
| | $L_K^c$ (GeV/c) | (5.6±0.3) | (18.4±1.7) | (76.0±8.8) |
| | $\chi^2/ndf$; Prob. | 6.863/10;0.7383 | 4.738/5;0.4486 | 4.522/6;0.6064 |
| | $a$(GeV$^{-2}$c$^3$) | (3.6±0.5)10$^8$ | (4.4±1.3) 10$^4$ | 8.8±3.2 |
| | $b$ (GeV/c)$^{-1}$ | 2.27±0.07 | 0.52±0.03 | 0.127±0.007 |
| 24 x10⁶ | $p_T^{min} \div p_T^{max}$ (GeV/c) | (1.05±0.05) ÷(4.4±0.4) | (6. ±0.4)÷(21.6±2.4) | (26.4±2.4) ÷(95.0±8.6) |
| | $J^c$ (GeV/c) | $J^c_{I-II}$ : (5.6±0.3) | $J^c_{II-III}$ : (24±1.7) | $J^c_{III-IV}$ : (100.0±8.6) |
| | $L^c_K$ (GeV/c) | (5.6±0.3) | (18.4±1.7) | (76.0±8.8) |
| | $\chi^2/ndf$; Prob. | 10.4/10;0.4061 | 4.22/5;0.5182 | 4.424/6;0.6194 |
| | $a$(GeV⁻²c³) | (1.4±0.2) 10¹¹ | (2.6±0.8) 10⁷ | 3705±1450 |
| | $b$ (GeV/c)⁻¹ | 2.23±0.07 | 0.53±0.04 | 0.127±0.008 |
| 25 x10⁸ | $p_T^{min} \div p_T^{max}$ (GeV/c) | (1.05±0.05) ÷(4.4±0.4) | (5.2±0.4)÷(16.8±2.4) | (21.6±2.4)÷(95.0±8.6) |
| | $J^c$ (GeV/c) | $J^c_{I-II}$ : (4.8±0.3) | $J^c_{II-III}$ : (19.2±1.7) | $J^c_{III-IV}$ : (100.0±8.6) |
| | $L^c_K$ (GeV/c) | (4.8±0.3) | (14.4±1.7) | (80.8±8.8) |
| | $\chi^2/ndf$; Prob. | 6.432/9;0.696 | 3.77/5;0.5829 | 7.511/7;0.3777 |
| | $a$(GeV⁻²c³) | (4.8±0.7)10¹³ | (3.5±1.2) 10¹⁰ | (2.5 ±0.9) 10⁶ |
| | $b$ (GeV/c)⁻¹ | 2.38±0.09 | 0.69±0.05 | 0.139±0.007 |
| 26 x10¹⁰ | $p_T^{min} \div p_T^{max}$ (GeV/c) | (1.05±0.05) ÷(4.4±0.4) | (5.2±0.4)÷(16.8±2.4) | (21.6±2.4)÷(95.0±8.6) |
| | $J^c$ (GeV/c) | $J^c_{I-II}$ : (4.8±0.3) | $J^c_{II-III}$ : (19.2±1.7) | $J^c_{III-IV}$ : (100.0±8.6) |
| | $L^c_K$ (GeV/c) | (4.8±0.3) | (14.4±1.7) | (80.8±8.8) |
| | $\chi^2/ndf$; Prob. | 8.671/9;0.4682 | 3.457/5;0.6299 | 7.469/7;0.3817 |
| | $a$(GeV⁻²c³) | (8.8±1.2) 10¹⁵ | (7.5± 2.6)10¹² | (3.9±1.8) 10⁸ |
| | $b$ (GeV/c)⁻¹ | 2.36±0.08 | 0.70±0.05 | 0.14±0.01 |

### A. Other evidence of the existence of the $p_T$ regions

As we have mentioned, article [6] told that the charged-particle $p_T$ spectrum contains, two quite distinct regions for the most central collisions: the exponential spectrum at $p_T \leq 4-5 \, \text{GeV}/c$ and power law behaviour at $p_T \geq 4-5 \, \text{GeV}/c$. These results could be evidence for existence of the $p_T$ regions. Furthermore, we would like to discuss following data from:

- PYTHIA simulation of the $p_T$ distributions for *pp* collisions [1, 5];
- the behaviour of the Nuclear Modification Factor as a function of the $p_T$ for Pb-Pb collisions [1].

### A.1. The PYTHIA simulation of the $p_T$ distributions

Figure 3a shows the ratio of the measured spectrum to the predictions of the four PYTHIA tunes [9÷10] to the interpolated spectrum of *pp* collisions at 2.76 TeV. The grey band corresponds to the statistical and systematic uncertainties of the measurement added in quadrature (the lower panel of the Figure 3a from the paper [1]). We have drawn 2 vertical lines only to show the extracted results of the fitting for the boundary values ($J^c$: for the first and second region - $J^c_{I-II}$; for the *II* and *III* region - $J^c_{II-III}$; for the *III* and *IV* region

- $J^c_{III-IV}$ ; for the *IV* and *V* region - $J^c_{IV-V}$, see Table I )[6] of the $p_T$ regions. One can see that the behavior of the ratios as a function of $p_T$ have some character: in different $p_T$ regions the simulation describes the experimental data differently. The PYTHIA D6T simulated spectrum shows that:
    - in the region of $p_T \leq 1\,\text{GeV}/c$ the value of the ratio is constant around ~1.2, then it decreases from ~ 1.2 to 1.0 in the region of $p_T = 1.0 - 3.4\,\text{GeV}/c$;
    - the ratio increases from ~1.0 to ~1.2 in the region of $3.4\,\text{GeV}/c < p_T \leq 6.0\,\text{GeV}/c$ and stays constant around ~ 1.2 in the region of $p_T = 6.0 - 22\,\text{GeV}/c$;
    - in the region of $p_T > 22\,\text{GeV}/c$ the ratio increases from ~1.2 till ~1.5.

These results indicate at least 3 $p_T$ regions for the behaviour of PYTHIA D6T ratio as a function of $p_T$: first, the region of $p_T < 3.4\,\text{GeV}/c$, then the region of $3.4 < p_T < 22\,\text{GeV}/c$, and, finally, the region of $p_T > 22\,\text{GeV}/c$. The boundary values of the regions are very close to extracted values from the experimental data. The values of the ratio for PYTHIA Perugia0 show 3 $p_T$ regions as well with boundary values of $p_T$ very close to ~3.4 GeV/c and ~22 GeV/c. PYTHIA 8 point out to existence of 3 regions but with boundary values of $p_T$: ~6.0 and ~22.0 GeV/c. Pythia ProQ20 shows 2 regions with boundary values of $p_T < 4 - 5\,\text{GeV}/c$. Here, we conclude that the comparisons of the experimental $p_T$ distributions for *pp* collisions with the predictions of four PYTHIA tunes at 2.76 A GeV indicate the existance of several regions with boundary values very close to ones extracted from the experimental data.

Figure 3b and 3c (taken from the lower panels a,b in Figure 5 of the paper [5]) show the ratio of the $p_T$ distribution measurement to the four PYTHIA tunes [5] for *pp* collisions at 0.9 and 7 TeV, respectively. We put two vertical lines on Figure 3b and three on Figure 3c to show the boundary values of the extracted $p_T$ regions from the experimental data. In case of *pp* collisions at 0.9 and 7 TeV the ratios as a function of $p_T$ have trends similar to ones observed in Figure 3a. Again, the simulation describes experimental data differently in different $p_T$ regions. For the *pp* collisions at 0.9 TeV (Figure 3b):

- in the first region, $p_T < 4 \text{ GeV}/c$, the ratio increases slowly for PYTHIA D6T predictions (from 1.1 to 1.2), speeds up for PYTHIA 8 ones from 1.0 to 1.4, decreases slowly for PYTHIA Perugia0 from 1.1 to 0.9 and flattens around ~ 1.0 for Pythia ProQ20 predictions;
- in the second region of $p_T \cong 4.0 - 17.2 \text{ GeV}/c$ the ratio increases for PYTHIA D6T predictions from 1.2 to 1.4, decreases slowly for PYTHIA 8 from 1.4 to ~ 1.2, stays constant at around ~0.8 for PYTHIA Perugia0 and reaches almost 1.0 for Pythia ProQ20 predictions ;
- in the third region with $p_T > 17.2 \text{ GeV}/c$ the ratio increases and flattens at following constants: ~ 1.6 for PYTHIA D6T, ~1.2-1.4 for PYTHIA 8 and around 1.0 for PYTHIA Perugia0 and Pythia ProQ20 (within experimental uncertainties).

---

[6] The values of $J^c$ were calculated using experimentally mesured values of $p_T^{min}$ (the value of $p_T^{min}(I)$ for regions *I* was taken as $p_T^{min}(I) = 0$) and $p_T^{max}$ (see Table I) as $J^c_{I-II} = (p_T^{max}(I) + p_T^{min}(II))/2$ for the region *I* and *II*, $J^c_{II-III} = (p_T^{max}(II) + p_T^{min}(III))/2$ for the region *II* and *III*; $J^c_{III-IV} = (p_T^{max}(III) + p_T^{min}(IV))/2$ for the region *III* and *IV* and $J^c_{IV-V} = (p_T^{max}(IV) + p_T^{min}(V))/2$ for the one *IV* and *V*. The variables $p_T^{min}(II), p_T^{min}(III), p_T^{min}(IV)$ and $p_T^{min}(V)$ represent the minimum values of transverse momentum in region *II,III,IV* and *V* respectively, whereas $p_T^{max}(I), p_T^{max}(II), p_T^{max}(III)$ and $p_T^{max}(IV)$ show the maximum values of transverse momentum in regions *I,II, III* and *IV* respectively. From the Table I one can see that we could get the value of $J^c_{II-III}$ (which is (101.1±7.1) GeV/c) for collisions with *c=4* only and in frame of uncertainties, this value is very close to the value of maximum mesured transverse momentum for this region: $p_T^{max}(III) = (91.2 \pm 10.0)\,\text{GeV}/c$. The table shows that the values of $p_T^{max}(III)$ in all cases are very close (in frame of uncertainties) to each other and around 100 GeV/c. So, we can fix the value of $p_T^{max}(III) \cong 100\,\text{GeV}/c$ for all collisions ( except the case where c=3) and to take the values of $J^c_{II-III}$ as $J^c_{II-III} = 100\,\text{GeV}/c$. The values of $J^c_{IV-V}$ was taken as $J^c_{IV-V} = p_T^{max}(IV) = (181 \pm 20)\,\text{GeV}/c$.

As we have mentioned above the *pp* collision data at 7 TeV shows four characteristically different regions of $p_T$ distributions. Fig.3c indicates that ratios of the measured $p_T$ distribution to ones from four PYTHIA models have different behaviours in the experimentally extracted regions. Various observations were made in following regions :

- for $p_T < 4.0 \text{ GeV}/c$ the ratio for PYTHIA D6T decreases from ~ 1.2 to 0.8, PYTHIA Perugia0 decreases from ~1.2 till ~0.8, PYTHIA ProQ20 decreases from ~ 1.2 to 0.8 and PYTHIA 8 increases from ~0.9 to ~1.0;
- for $4 \text{ GeV}/c < p_T < 17.2 \text{ GeV}/c$ the ratio for PYTHIA D6T increases from 0.8 to 1.0, PYTHIA Perugia0 decreases and stays constant at 0.7, PYTHIA ProQ20 increases from 0.8 to 0.9 and PYTHIA 8 it is constant at around ~1.0;
- for $17.2 \text{ GeV}/c < p_T < 105.5 \text{ GeV}/c$ the ratio for PYTHIA D6T stays almost constant at 1.2, PYTHIA Perugia0 at 0.8, PYTHIA ProQ20 at 1.0 and PYTHIA 8 at around 1.1;
- for $p_T > 105.5 \text{ GeV}/c$, again, the ratio for PYTHIA D6T stays constant at 1.2, PYTHIA Perugia0 at 0.8, PYTHIA ProQ20 and PYTHIA 8 stay constant at 1.0.
-

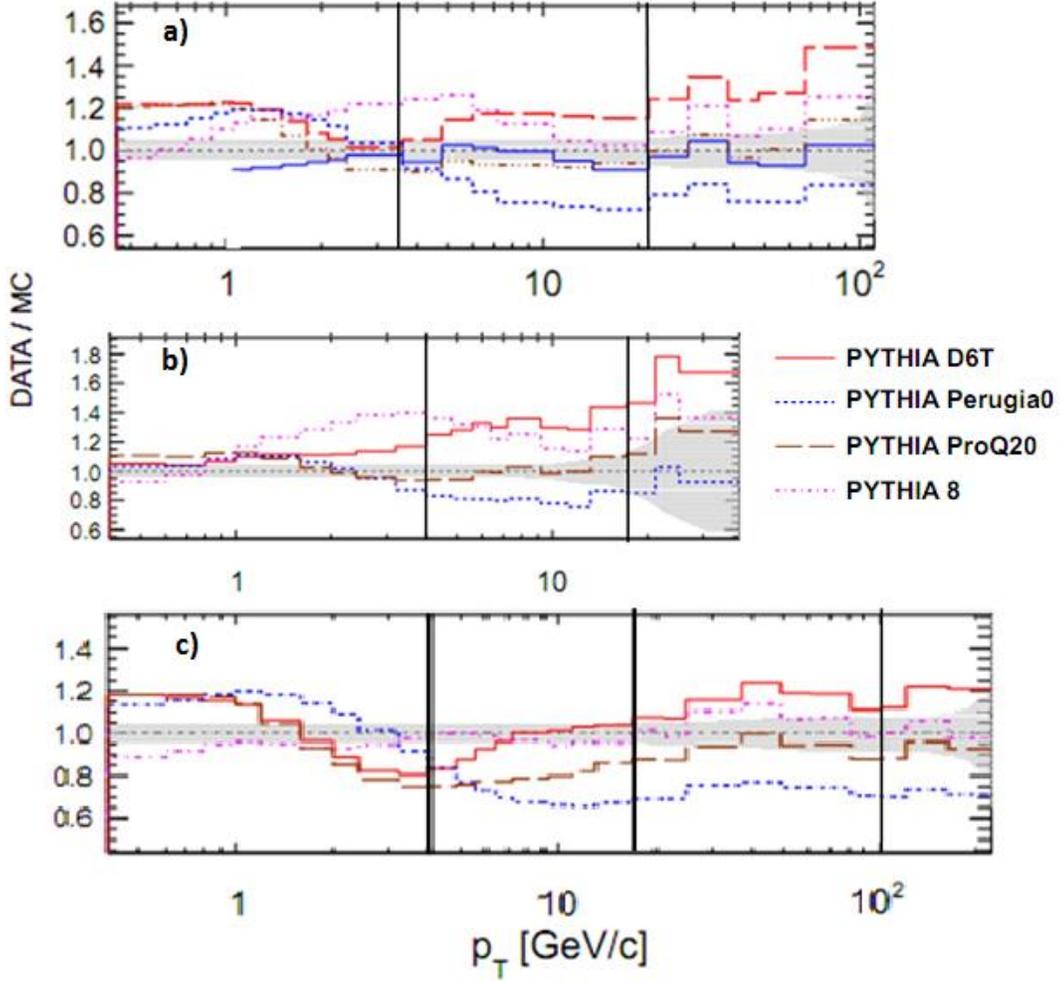

FIG. 3a-c. The ratio of the measured spectrum for *pp* collisions at: a) $\sqrt{s}=2.76$ TeV (lower panel of the Figure 3a from paper [1]) ; b) $\sqrt{s}=0.9$ TeV (lower panel in Figure 5a from the paper [5]); c) $\sqrt{s}=7$ TeV (lower panel in Figure 5b from the paper [5]) to the predictions of the four PYTHIA tunes and to the interpolated spectrum (in Figure 3a). The grey band corresponds to the statistical and systematic uncertainties of the measurement added in quadrature.

We conclude that the above observations about the ratios of measured $p_T$ spectra to the simulated PYTHIA models for the *pp* collisions at 0.9 TeV, 2.76 TeV and 7 TeV indicate the existence of several different $p_T$ regions (except maybe the PYTHIA ProQ20 predictions for pp collisions at 0.9 TeV) for the behaviour of the invariant charged particle differential yield. All four PYTHIA models give similar predictions for the $p_T$ regions *III* and *IV* for *pp* collisions at 7 TeV.

## A.2. The behaviour of the Nuclear Modification Factor as a function of $p_T$

Figure 4 shows a behaviour of the $R_{AA}$ for the charged primary particles (with pseodourapidity $|\eta|<1$) produced in Pb-Pb collisions at 2.76 A TeV [1] as a function of $p_T$ in six centrality bins (Figure 5 from [1]). We put four vertical lines to show extracted $p_T$ regions as a result of the fitting (two black lines for Pb-Pb collisions and two red ones for *pp* collisions) on Figure 4. There are at least three regions for the behaviour of the $R_{AA}$ values as a function of $p_T$: the first region is $p_T < 4-5\,\text{GeV}/c$, the second region is $4-5\,\text{GeV}/c < p_T < 20\,\text{GeV}/c$ and the third one is $p_T > 20\,\text{GeV}/c$. The boundary values of $p_T$ for different regions are very close to ones acquired form fitting the $p_T$ distributions. So, we conclude that the behaviour of the $R_{AA}$ values as a function of $p_T$ suggests the existence of at least three regions. It is important to note that since *pp* and Pb-Pb collisions have different boundary values for the $p_T$ regions, the behaviour of $R_{AA} = f(p_T)$ gets more complex. It could be made cleaner if this difference was taken into account during the construction of the $R_{AA}$.

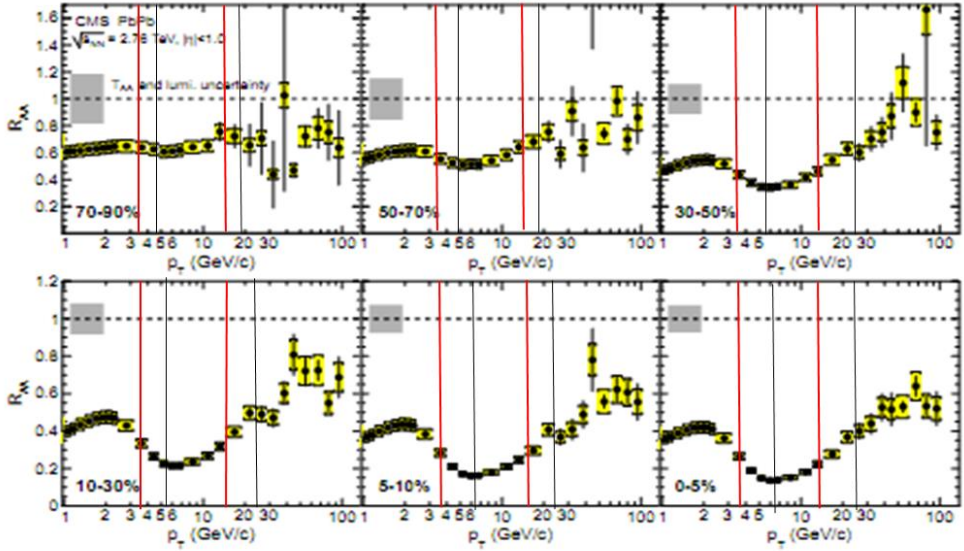

**Fig. 4** $R_{AA}$ values as a function of $p_T$ for six Pb-Pb centrality bins at 2.76 A TeV [1]. The error bars represent statistical uncertainties and yellow boxes represent the $p_T$-dependent systematic uncertainties.

### III. STUDY OF THE CHARACTERISTICS OF $p_T$ REGIONS

## A. The length of the regions

Using the data at boundary values of $p_T$ regions ($J^c$, see Table I) we defined the lengths for the different $p_T$ regions $K=I,II,III,IV$ ($K$ is the number of region) in various collisions $c$ ($L_K^c$). The values of $L_K^c$ were calculated as $L_I^c = J_{I-II}^c$; $L_{II}^c = J_{II-III}^c - J_{I-II}^c$; $L_{III}^c = J_{III-IV}^c - J_{II-III}^c$ and $L_{IV}^c = J_{IV-V}^c - J_{III-IV}^c$ (see Table I). One can see from the data in Table I that:

$$L_I^c < L_{II}^c < L_{III}^c \quad \text{for} \quad c=1, 21\text{-}26, \qquad (1)$$

$$L_{IV}^4 < L_{III}^4, \qquad (2)$$

$$L_{III}^3 < L_{II}^3 \qquad (3)$$

The result (1) shows that the lengths of $p_T$ regions increase in direction of the growing $p_T$ (or with $K$). The results (2) and (3) indicate that the number of experimental points is not enough to define the length of the region $III$ in case of $c=3$ and the length of region $IV$ in case of $c=4$.

For the $pp$ collisions one can see that:

$$L_I^1 \cong L_I^3 \cong L_I^4, \qquad (4a)$$

$$L_{II}^1 > L_{II}^3 \cong L_{II}^4, \qquad (4b)$$

$$L_{III}^1 \cong L_{III}^4 > L_{III}^3, \qquad (4c)$$

$$L_{III}^4 > L_{IV}^4 \qquad (4d)$$

The result (4a) indicates that the length of region $I$ for $pp$ collisions doesn't depend on incident energy and $|\eta|$ interval. Then the result (4b) means that in region $II$ of $pp$ collisions the length doesn't depend on energy but depends very weakly on $|\eta|$. In region $III$ the length (as indicated in (4c)) for $c=3$ is less than length for the collisions with $c=1$ and $4$. As it was mentioned above, this is due to impossibility to see the full length of region $III$ in case of $c=3$ (during the fitting $ndf$ for this case was 1 only). The result (4d) again due to impossibility to see the full length of region $IV$ in case of $c=4$ (again the $ndf = 1$). Here we can conclude that in region $III$ the lengths don't depend on incident energy and $|\eta|$. The data from Table I show:

$$L_I^{21\div22} > L_I^{23\div24} > L_I^{25\div26} > L_I^1 \cong L_I^3 \cong L_I^4, \qquad (5)$$

$$L_{II}^{1,21\div26} = [(14.4\pm1.7) \div (18,4\pm1.2)]\,\text{GeV/c}, \quad L_{II}^{3,4} = (13.2\pm1.4)\,\text{GeV/c} \quad (6)$$

$$L_{III}^c = (76.6\pm8.8) \div (80.8\pm8.8)\,\text{GeV/c} \qquad (7)$$

The result (5) demonstrates that the length of region $I$ depends on the mass of projectiles, centrality, incident energy and $|\eta|$ in the Pb-Pb collisions. We observe 4 groups: $c=21$-$22$, $c=23$-$24$, $c=25$-$26$ and $c=1,3$-$4$. The length of groups seem to decrease from the central events to peripheral ones, and they are the smallest for $pp$ collisions. The length of region $II$ almost doesn't depend on the mass of projectiles and centrality of collisions but depends weakly on the energy of collisions and $|\eta|$ (result (6)). The length of region *III* doesn't depend on the mass of the projectiles, their centrality, energy and $|\eta|$ (from result (7)).

## B. The behavior of the values of the parameter *b* as a function of $L_K^c$

For easy viewing of the values of the parameters *b* (from Table I) in different $p_T$ regions and for different collisions *c*, we presented them as a function of $L_K^c$ in Figure 5. This distribution was fitted with the function of $y' = a'e^{-b'x'}$ (*a'* and *b'* are the free fitting parameters). It can be seen that the fitting selects two regions in $L_K^c$ with the following results: $\chi^2 = 14.17/17$, prob.= 0.6552, $a' = 4.5 \pm 0.2$, $b' = 0.122 \pm 0.006$ for the first region and $\chi^2 = 5.875/7$, prob.= 0.5544, $a' = 0.4 \pm 0.1$, $b' = 0.014 \pm 0.004$ for the second region. In the case of c=3 for *III* region, the value of $b'$ is located between two regions with a large error and with $\chi^2/ndf = 0.0005464/1$. In order to accurately determine the position of this point, either new measurements in the region of $p_T > 30 \text{ GeV}/c$ or an increase in the number of measured points in the region of $p_T > 20 \text{ GeV}/c$ are required. The same problem with the point in the case of the region *IV* for *pp* collisions at 7 TeV, here $\chi^2/ndf = 0.007805/1$ and again to determine exact position of the point, it is necessary to make either new measurements in the region of $p_T > 180 \text{ GeV}/c$ or increases of the number of measured points in the region of $p_T = 110 - 180 \text{ GeV}/c$. Therefore, we will not discuss the positions of these two points. Of the other observations from Figure 5 and from Table I are:

(1) in the region *I*: $b^1 \cong b^3 \cong b^4 > b^{21} \cong b^{22} \cong b^{23} \cong b^{24} \cong b^{25} \cong b^{26}$, where the superscript indicates different collisions *c*. The values of *b* "slip" dependent on $L_K^c$ and grouped depending on *c*, show the existence of four groups: the first group combines events with *c* = 1,3-4; second group unites the occurrences with *c* = 21-22; third group includes the case of *c* = 23-24; the fourth group combines collisions with *c* = 25-26. It is interesting that the "slip" is directed from the minimum value of $L_K^c$ for collision with *c* = 25-26 toward the peripheral Pb-Pb collisions then the semi central Pb-Pb collisions (*c* = 23-24) and to the maximum value of $L_K^c$, for central Pb-Pb collision (*c*= 21-22). Thus, the groups show the dependence of the parameter *b* on centrality. We have already noted on top that, the lengths for the first regions are also grouped into 4 groups of *c*: *c* = 21-22; *c* = 23-24; *c* = 25-26 ; *c* = 1,3-4 and the lengths of the groups are decreased in the direction from the central to peripheral, so they would be minimum for the *pp* collisions. Thus, we can observe a likeness between the length of the region and the value of the parameter *b* in this group which is understandable because the parameter *b* has the meaning of the inverse of $p_T$ ;

(2) in the *II* $p_T$ region, the values of the parameters *b* show: $b^1 \cong b^3 \cong b^4 \cong b^{25} \cong b^{26} > b^{21} \cong b^{22} \cong b^{23} \cong b^{24}$. There is a certain weak "slip" (weaker than in the *I* region) for the values of *b* and 2 groups by *c*: the first group is for *c* = 1,3-4, 23-26, and the second group for $c = 21$-$22$. The "slip" has a direction of $c = 1,3$-$4, 23$-$26 \rightarrow c = 21$-$22$. It was mentioned above that the length of the *II* region does not depend on the mass of projectiles and the centrality of collisions, but it is slightly dependent on the primary energy and $|\eta|$. In this way, one can see again the same likenesses between $L_K^c$ and the parameter *b*.

(3) in the third region: $b^1 \cong b^3 \cong b^4 \cong b^{21} \cong b^{22} > b^{23} \cong b^{24} \cong b^{25} \cong b^{26}$, and one cannot see any "slip" and only one group. The values of the $L_K^c$ for this region does not depend on the characteristics of the collisions. There is a $L_K^c$-*b* likenesses again.

The parameter *b* as a function of the length of the region shows that the *I* and *II* $p_T$ regions are connected to each other. The *III* and *IV* regions are also seem to be connected to each other, but the measurement errors in the *IV* region are large, and *ndf* = 1 which does not allow to reach a conclusion for any connection for these regions.

Thus, the important result is that in the first $p_T$ region, the values of *b* strongly depend on the characteristics of the interaction in the second region, the dependence weakens and it is completely absent in the third region.

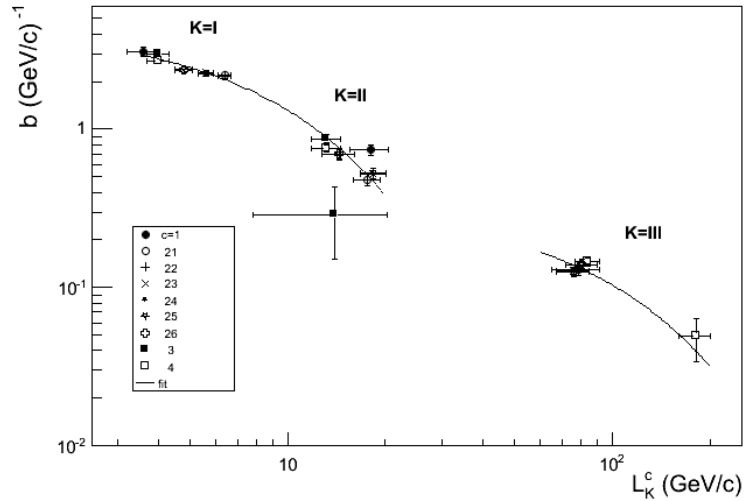

FIG.5. The values of the parameter *b* as a functin of $L_K^c$.

## C. The behavior of the parameter *a* as a function of $L_K^c$

Figure 6a shows the behavior of the second fitting parameter[7] $a^c$ as a function of $L_K^c$. It can be observed that:

(1a) in the I region there is: $a^1 \cong a^3 \cong a^4 < a^{21} < a^{22} < a^{23} < a^{24} < a^{25} \cong a^{26}$, where the superscripts indicate $c$. The values of $a$ "slip" dependent on $L_K^c$ and they are grouped into 4 groups with: $c = 1,3\text{-}4$; $c = 25\text{-}26$; $c = 23\text{-}24$ and $c = 22\text{-}21$. To clarify the picture, 4 lines are drawn on it, which combine points from different groups and quantitatively characterize the "slippage". The "slip" has a direction: from the minimum value of the length for $pp$ collisions to $c = 25\text{-}26$, the peripheral Pb-Pb collisions, then the semi central Pb-Pb collisions ($c = 23\text{-}24$) and to the maximum value of $L_K^c$, for central Pb-Pb collisions ($c = 21\text{-}22$) with step $\delta L_c \cong 1$. But within the groups there are no differences for the values of $a$;

(2a) in the second region, it can be observed that: $a^1 \cong a^3 \cong a^4 < a^{21} < a^{22} < a^{23} < a^{24} < a^{25} < a^{26}$. There is some weak (in comparison with the region I) "slip" for the values of $a$ as a function of $L_K^c$ and there exist two groups: the first group for the collisions with $c = 1,3\text{-}4, 25\text{-}26$; the second one for the collisions with $c = 21\text{-}24$. The "slip" is directed from the first group to the second one. The values of the parameter $a$ are decreased from the peripheral Pb-Pb collisions to the most central ones, and reach a minimum for the $pp$ collisions.

(3a) it can be seen that, in the III region: $a^1 < a^3 < a^4 < a^{21} < a^{22} < a^{23} < a^{24} < a^{25} < a^{26}$, and there is only one group, without any "slip". Again, the values of the parameter $a$ are decreased from the peripheral interactions to the most central ones, and reach a minimum for the $pp$ collisions.

Figure 7a presents the dependence of the ratios for the parameters $a$ on the length for Pb-Pb collisions with different values of the centrality ($a^c$) to the values of the parameter $a$ in the 3 $p_T$-regions for $pp$ collisions ($a^1$). It can be observed that the behaviors of $a^c/a^1$ are different for peripheral and central collisions. For peripheral collisions, the ratios are increased almost linearly with the lengths, though they are significantly different for the semi-central collisions. The last mainly because of the point for the second region, it "fails" and the "fails" get "deeper" with centrality and reaches a maximum value for the most central collisions. Figure 7b shows the same dependencies of the ratios for the parameters $b$. It is seen that within the uncertainties, this ratio is increased almost linearly with lengths. There is no any "fails". The observed "fails" can be a signal about anomalous suppression of high $p_T$ particles in the region of $p_T = 5 - 20 \text{ GeV}/c$.

Figure 8a,b demonstrate the ratios of $R_1^{'} = a^c/(a^1 <N_{part}>)$ (Figure 8a) and $R_2^{'} = a^c/(a^1 <N_{coll}>)$ (Figure 8b) as a function of $L_K^c$ in which $a^1$, $<N_{part}>$ and $<N_{coll}>$ are respectively the values of the parameter $a$ for $pp$ collisions at 2.76 TeV, the average number of participants' nucleons, and the number of binary nucleon-nucleon collisions (obtained from Table 1 in [1]). Figure 8a and 8b show that, the behaviors of $R_1^{'}$ and $R_2^{'}$ are almost same and

---

[7] The values of $a$ were divided into: $10^2$ for the collisions with $c=22$; $10^4$ for the events with $c=23$; $10^6$ for the cases with $c=24$; $10^8$ for the collisions with $c=25$; $10^{10}$ for the events with $c=26$, because they were multiplied by these numbers in the paper [1] from which the data were taken.

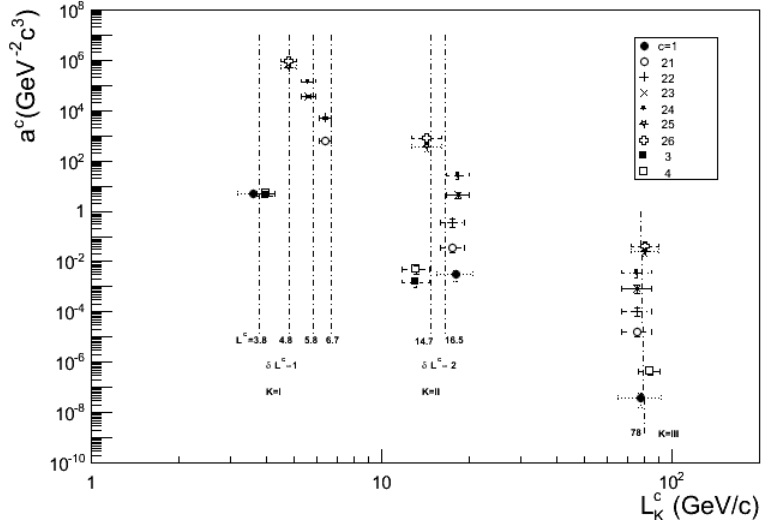

FIG. 6. The values of the parameter $a^c$ as a function of the lengths. The $\delta L_c$ indicates the slips inside the groups.

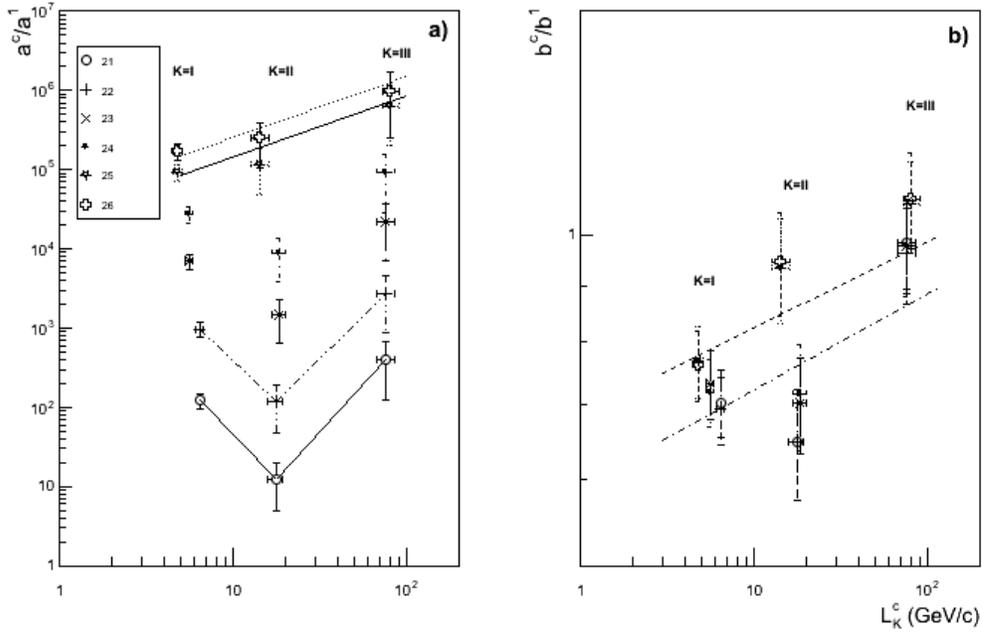

FIG.7. $L_K^c$ – dependences for the ratios of: a) the $a^c/a^1$  b) the $b^c/b^1$.

depend on the centrality. The central collisions have a minimum value for the $R_1^{'}$ and $R_2^{'}$ in the regions *II*. It can be seen that there is some kind of suppression for the most central Pb-Pb collisions ($c = 21$) with the values of $R_1^{'}$ and $R_2^{'} < 1$ and this suppression is much stronger for the *II* region compared to the *I* and *III* region.

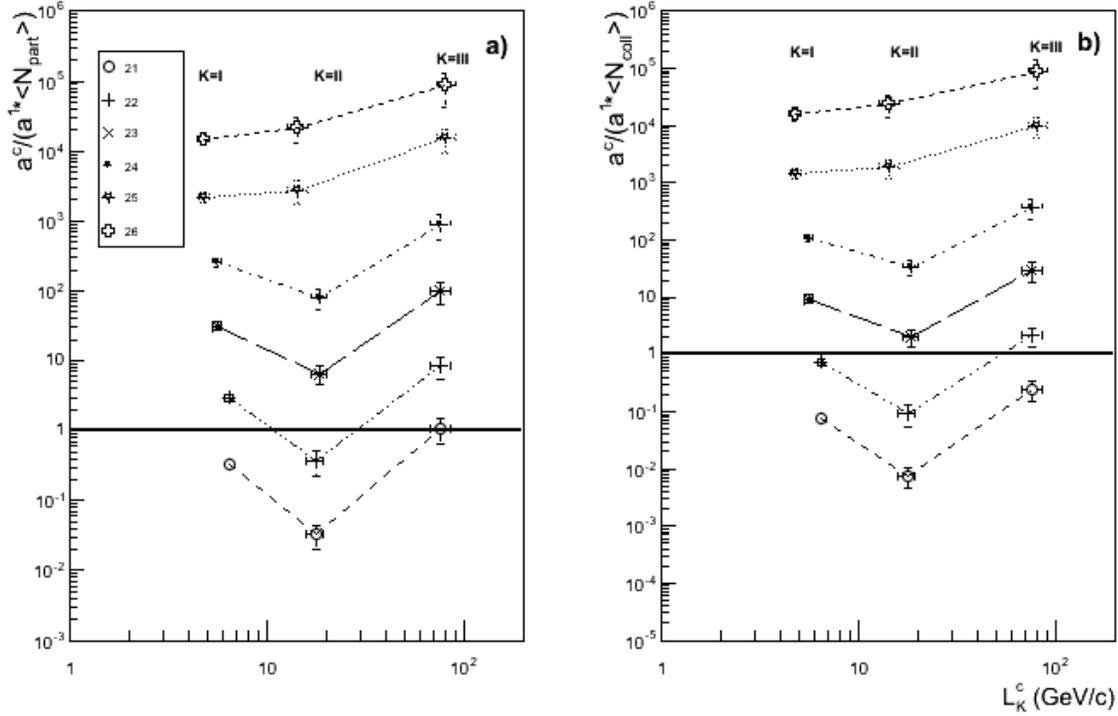

FIG. 8. The values of the: a) $a^c/(a^1 <N_{part}>)$ ; b) $a^c/(a^1 <N_{call}>)$ as a function of $L_K^c$.

In conclusion, the results of (1a) ÷(3a) indicate that the values of the parameter $a$ depend on:

- 2 variables: $L_K^c$ and centrality in the first and second regions;
- only one variable - centrality in the *III* region.

Another problem is that whether if the second region can be considered as an independent one or there are two independent regions *I* and *III* and the region *II* is the result of summing the regions. To answer this question, we constructed Figure 9 using the interpolation data for *pp* collisions at 2.76 TeV. The function of $y_I = 5.16e^{-3.1p_T}$ (the values of 5.16 and 3.1 are the best fitting results in the first region for the case of $c = 1$, which were taken from Table I without errors); $y_{II} = 0.003e^{-0.74p_T}$ (the values of 0.003 and 0.74 are the best results of fitting in the second region for the case of $c = 1$, taken from Table I without errors); $y_{III} = 410^{-8}e^{-0.13p_T}$ (the values of 4 $10^{-8}$ and 0.13 are the best fitting results in the third region for the case of $c = 1$ that were taken from Table I, without errors); $y_I + y_{II}$ and $y_I + y_{II} + y_{III}$ were used for interpolation. The behavior of the functions in Figure 9 demonstrates that the *II* region in the *pp* collisions at 2.76 TeV cannot be considered as a simple sum of the *I* and *III* regions, this is a more complex region. As mentioned above, the dependence of the fitting parameters' values in this region is also different from those that were in the first and the second regions. This also confirms that the *II* $p_T$ region is a region with special properties.

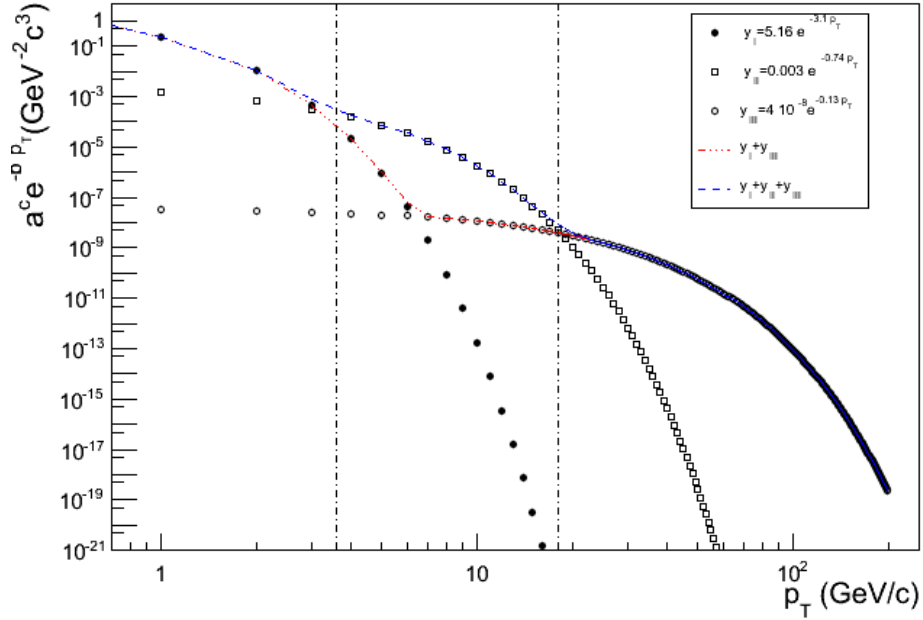

FIG. 9. The $p_T$ distributions for the $a^{c=1}e^{-b^{c=1}p_T}$ in: the *I* region $y_I$ ; the *II* region $y_{II}$ ; the *III* region $y_{III}$. The red line is a sum of $y_I + y_{II}$ and the blue one does $y_I + y_{II} + y_{III}$. The vertical lines show the boundary values of the detected $p_T$ regions.

## IV. MAIN RESULTS

It was observed that the experimentally measured $p_T$ distributions of the invariant differential yield for charged primary particles produced in *pp* collisions at 0.9 TeV, 2.76 TeV and 7 TeV, and in Pb-Pb collisions with 6 centrally bins at energy of 2.76 A TeV contain at least 3 different $p_T$ regions. In frame of the regions the behavior of the distributions obeys the exponential law of $ae^{-bp_T}$. The regions have special properties. The second region cannot be considered as a simple sum of *I* and *III* regions, this is a complex region. The behavior of the parameter *b* as a function of the length indicates that the *I* and *II* $p_T$ regions are connected to each other, also the *III* and *IV* regions seem to be connected to each other. The lengths of the regions increase with $p_T$.

For *pp* collisions, there were not observed any essential dependences for the fitting parameters on the characteristics of the collisions (a certain $|\eta|$-dependence in the *II* region could be an exception). While for Pb-Pb collisions, it was obtained that the values of the fitting parameters strongly depend on the characteristics of the collisions (medium effect) in the region of $p_T < 4-6\,\text{GeV}/c$ (*I* region). The dependences become weaker in the second region of $p_T$ (

$4-6\,\text{GeV}/\text{c} < p_T < 17-20\,\text{GeV}/\text{c}$) and almost disappeared in the *III* region of $p_T > 17-20\,\text{GeV}/\text{c}$ (except centrality dependences for the parameter *a*).

In the case of central collisions, the ratio of the $a^{21-24}$ values for Pb-Pb collisions to the values of $a^1$ for *pp* collision multiplied by the average number of participant nucleons (or the number of binary nucleon-nucleon collisions) have the minimum values for the second region and, it becomes less than 1 for the most central events ($a^{21-22}$) – a suppression, and maximum suppression is observed in the *II* region. This effect is a result of the $a^c/a^1$-ratio "fail" in the *II* region.

## V. THE DISCUSSION OF THE RESULTS

Observation of the $p_T$ regions for both *pp* and for Pb-Pb collisions with different centralities suggests that these regions reflect the fragmentation and hadronization properties of partons.

The parameter *b* can be represented as $b = \dfrac{1}{\bar{p}_T}$ (in which $\bar{p}_T$ is some average $p_T$ for parton systems), and with considering $-(\bar{p}_T) \cong t = -q^2$ (*t* is the Mandelstam variable), we can write that $b \cong \dfrac{1}{\sqrt{q^2}}$ and $q \cong 1/b^2$. Table II shows the values of $q^2$ and $\alpha_s \cong \left[\ln(q^2/\Lambda^2)\right]^{-1}$ at $\Lambda = 0.2\,\text{GeV}/\text{c}$. The results show that the values of $\alpha_s$ don't depend on the primary energy, the mass of projectiles and centrality in the region *III*. In this third region, they are less than unit and could indicate that parton-parton collisions are dominant in this region, and there is not any essential medium effect to modify of partons. This can be resulted by the effect of nuclear transparency [11]. In the second region, the values of $\alpha_s$ are increased more than 1.5 times staying less than one, while there is a certain weak dependence on the interaction parameters. In the *I* region the values of $\alpha_s$ are increased ~4÷6 times (compared to the *III* region) and reach the values of $\alpha_s \sim 1$ for the *pp* collisions. Some strong dependences for the values of $\alpha_s$ on collisions' characteristics were observed in the *I* region. The discussed increase in the value of $\alpha_s$ during the transition from the third region to the second, and then to the first region, is similar to the dependence of $\alpha_s$ on $p_T$, which is characteristic of the QCD quark string: $\dfrac{1}{r^2} \sim Q^2 = -q^2$ ( in which *r* is a distance between quarks in the string). So it can be stated that the parameter *b* could characterize the distance among quarks in a string.

As it mentioned above, the lengths of regions are increased with $p_T$. Table III presents the ratio values of lengths in different $p_T$ regions. It can be observed that for Pb-Pb collisions, $L^c_{III} : L^c_{II} \cong 5$ and $L^c_{II} : L^c_{I} \cong 3$ for all cases. For *pp* collisions at energy of 2.76 TeV, the ratios were $L^1_{III} : L^1_{II} \cong 5$ and $L^1_{II} : L^1_{I} \cong 5$. This result can serve as an additional clue that the fragmentation and hadronization of partons occurs through the string dynamics, and the values of $L^c_K$ can be related to the string tension [12]. Strings of the first generation as the most energetic (from the region *III*)[8] ones can decay into strings of the second generation (region *II*) and form high $p_T$ partons / hadrons in the third region (perhaps this region is the best one to study the string dynamics). In its turn, strings of the second generation can decay into new strings of the third generation (region *I*) and form high $p_T$ partons / hadrons in the second region. As mentioned in the case of Pb-Pb collisions, the ratio is $L_{II}^c : L_I^c \cong 3$, but the values of the ratio are about 5 for *pp* collisions. This result could indicate that due to medium effect in Pb-Pb collisions in the second region, there is energy loss by the strings. The strings of the second generation can decay and produce strings / partons / hadrons of the region *I*.

---

[8] Observation of the fourth region in the case of *c* = 4 indicates that region *III* is not the region of the first generation of partons for *c* =1,21-26. More regions could appear with more $p_T$ measurements for $p_T > 100\,\text{GeV}/\text{c}$.

The increase in the density of strings in the second region can be reason of the fusion of strings [13]. The behavior of $a^c / (a^1 <N_{part}>)$ (or $a^c / (a^1 <N_{call}>)$ ) changes and the ratio in the second region becomes minimal, and anomalous suppression is observed for the most central Pb-Pb collisions. We assume that this can be due to the collective effects associated with the fusion of strings [13] and the appearance of a new string in the second region for the most central events.

In this approach the first $p_T$ region is a region of maximum number of hadrons (minimum number of strings) because of the directly hadronization of the low energy strings to mesons - "Cronin enhancement" [14]. Lets examinine the RHIC data [15] (see Figure 10) on the behavior of the invariant yield of identified particles emitted near midrapidity as a function of $p_T$ in central Au-Au collisions at energy 200 GeV per nucleon .The fact that the yield of 2 quarks systems ($\pi^\pm$ - and $K^\pm$) is much greater than 3 quarks systems (*p* and *anti p*) at low $p_T$ could be a proof of direct hadronization of the low energy strings into mesons from this region.

TABLE II. The values of the $q^2 (GeV/c)^2$ and $\alpha_s$.

| c | $q^2 (GeV/c)^2$ | | | $\alpha_s$ | | |
|---|---|---|---|---|---|---|
| | I | II | III | I | II | III |
| 3 | 0.111 ±0.007 | 1.3±0.2 | - | 0.98±0.06 | 0.286±0.009 | - |
| 4 | 0.14±0.01 | 1.7 ±0.2 | 47.2 ±0.5 | 0.81±0.05 | 0.265±0.007 | 0.141±0.0002 |
| 1 | 0.10±0.01 | 1.8±0.3 | 59.2± 9.0 | 1.1± 0.1 | 0.26±0.01 | 0.137±0.003 |
| 21 | 0.21±0.01 | 4.3±0.7 | 61.0±6.7 | 0.60± 0.02 | 0.213±0.008 | 0.136 ±0.002 |
| 22 | 0.22±0.01 | 4.3 ±0.7 | 64.0±8.2 | 0.59± 0.02 | 0.213±0.008 | 0.136±0.002 |
| 23 | 0.19±0.01 | 3.7±0.4 | 62.0± 6.8 | 0.63±0.03 | 0.221±0.006 | 0.136 ±0.002 |
| 24 | 0.20±0.01 | 3.6±0.5 | 62.0±7.8 | 0.62±0.02 | 0.223±0.007 | 0.136 ±0.002 |
| 25 | 0.18±0.01 | 2.1±0.3 | 51.8±5.2 | 0.67±0.03 | 0.253±0.009 | 0.140±0.002 |
| 26 | 0.18±0.01 | 2.0±0.3 | 51.0±7.3 | 0.67±0.03 | 0.254±0.009 | 0.140±0.003 |

TABLE III. The ratios for the regions' lengths.

| C | $L_{III}^c : L_{II}^c$ | $L_{II}^c : L_I^c$ |
|---|---|---|
| 1 | (4.4±0.8) | (5.0±0.7) |
| 21 | (4.3±0.7) | (2.8±0.3) |
| 22 | (4.3±0.7) | (2.8±0.3) |
| 23 | (4.1±0.6) | (3.3±0.4) |
| 24 | (4.1±0.6) | (3.3±0.4) |
| 25 | (5.6±0.9) | (3.0±0.4) |
| 26 | (5.6±0.9) | (3.0±0.4) |

## VI. CONCLUSION

We have undertaken an analysis of the data and concluded the following explanation for observed $p_T$ regions at the LHC energies: 1) region *III* (the region of $p_T > 17-20\,\text{GeV/c}$) is the domain of creation of first-generation partons/strings during collisions, where the most energetic hadrons / partons / strings (with highest tension) are produced and weakly modified by the medium due to the effect of nuclear transparency; 2) region *II* ($4-6\,\text{GeV/c} < p_T < 17-20\,\text{GeV/c}$) is the one with highest density of the strings decayed from ones in region *III*. The high density causes string fusion and a collective phenomenon, which is as a result of the new string formation in the most central Pb-Pb interactions. This region could be responsible for the Quark Gluon Plasma formation through the fusion of strings and explain anomalous behaviour of the Nuclear Modification Factor; 3) The first $p_T$ region ($p_T < 4-6\,\text{GeV/c}$) is the one with the maximum number of hadrons and minimum number of strings due to direct hadronization of the low energy strings into two quark systems - mesons

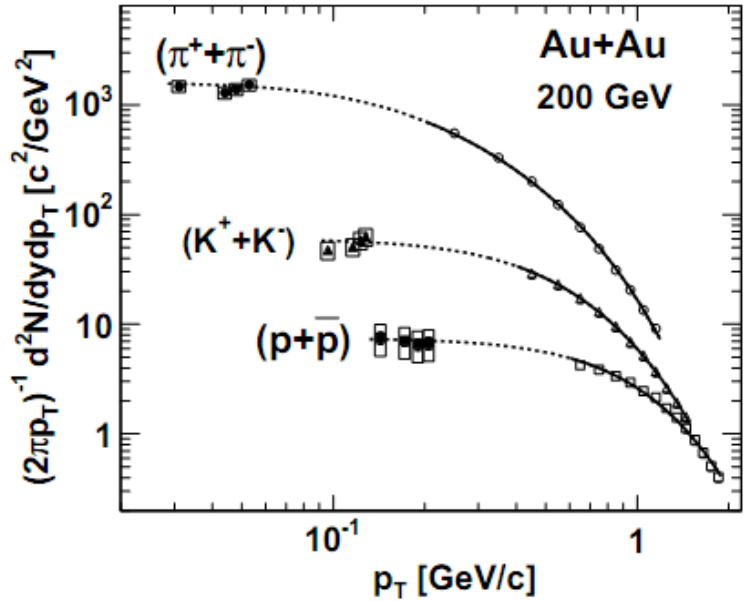

FIG. 10. The invariant yield of identified particles emitted near midrapidity as a function of $p_T$ in central Au-Au collisions at energy 200 GeV per nucleon from PHENIX at higher momenta [16] and PHOBOS at lower momenta [17] (the figure was taken from the paper[15])

The observation of $p_T$ regions for both *pp* and Pb-Pb collisions may point towards a similarity of particle production in these events. Several effects typical for heavy-ion phenomenology have been observed in high-multiplicity *pp* collisions [18]. These similarities can be understood qualitatively by string fusion: with growing of the density of strings, the multiplicity of particles increases and when the string density reaches a critical value (high multiplicity) similar to those observed in heavy-ion collisions, then they start to fuse, forming a new string.

## Acknowledgement

I would like to acknowledge the CIIT Islamabad Pakistan, which provided suitable platform and all possible facilities to perform the analysis and Aziza Suleymanzade (PhD Candidate Simon and Schuster Labs of University of Chicago) for her essential help during preparing the text.

___________________________________________